\documentclass[a4paper,12pt]{article}
\usepackage[english]{babel}
\usepackage{jheppub2}
\usepackage{subfig}
\usepackage[T1]{fontenc} 
\usepackage{graphicx}
\usepackage{epsfig}
\usepackage{amsmath}
\usepackage{amssymb}
\usepackage{float}
\usepackage{placeins}
\usepackage{braket}
\usepackage{slashed}
\usepackage{mathdots}
\usepackage{lipsum}

\allowdisplaybreaks[4]

\title{Operator mixing, UV asymptotics of nonplanar/planar $2$-point correlators, and nonperturbative large-$N$ expansion of QCD-like theories}

\author[a]{Ugo Aglietti}
\author[b]{Matteo Becchetti}
\author[c]{Marco Bochicchio}
\author[a,c]{Mauro Papinutto}
\author[a,c]{Francesco Scardino}
\affiliation[a]{Physics Department, Sapienza University,\\Piazzale A. Moro 2, Roma, I-00185, Italy}
\affiliation[b]{Physics Department, Torino University and INFN Torino, \\
Via Pietro Giuria 1, I-10125 Torino, Italy}
\affiliation[c]{Physics Department, INFN Roma1, \\
Piazzale A. Moro 2, Roma, I-00185, Italy}
\emailAdd{ugo.aglietti@roma1.infn.it}
\emailAdd{matteo.becchetti@unito.it}
\emailAdd{marco.bochicchio@roma1.infn.it}
\emailAdd{mauro.papinutto@roma1.infn.it}
\emailAdd{francesco.scardino@roma1.infn.it}

\abstract{We work out the interplay between lowest-order perturbative computations in the 't Hooft coupling, $g^2=g^2_{YM} N$, operator mixing, renormalization-group (RG) improved ultraviolet (UV) asymptotics of leading-order (LO) nonplanar/planar contributions to $2$-point correlators, and nonperturbative large-$N$ expansion of perturbatively massless QCD-like theories. As concrete examples, we compute to the lowest perturbative order in $SU(N)$ YM theory the ratios, $r_i$, of LO-nonplanar to planar contributions to the $2$-point correlators in the orthogonal basis in the coordinate representation of the gauge-invariant dimension-$8$ scalar operators and all the twist-$2$ operators.
The very definition of $r_i$ depends on the existence of the aforementioned basis, in such a way that its meaning is apparently limited to the lowest perturbative order only. Yet, we demonstrate that -- if $\frac{\gamma_0}{\beta_0}$ has no LO-nonplanar contribution, with  $\gamma_0$ and $\beta_0$ the one-loop coefficients of the anomalous-dimension matrix and beta function respectively -- $r_i$ actually coincides with the corresponding ratio in the large-$N$ expansion of the RG-improved UV asymptotics of the $2$-point correlators, provided that a certain canonical nonresonant diagonal renormalization scheme exists for the corresponding operators. Contrary to the aforementioned scalar operators, for the first $10^3$ twist-$2$ operators we actually verify the above conditions, and we get the universal value $r_i=r_{twist-2}=-\frac{1}{N^2}$. Hence, nonperturbatively such $r_i$ must coincide with the UV asymptotics of the ratio of the glueball self-energy loop to the glueball tree contribution to the $2$-point correlators above. As a consequence, the universality of $r_{twist-2}$ reflects the universality of the effective coupling in the nonperturbative large-$N$ YM theory for the twist-$2$ operators in the coordinate representation.}

\DeclareMathOperator{\Tr}{Tr}
\newcommand{\beq}{\begin{equation}}
\newcommand{\be}{\begin{equation}
\newcommand{\ee}{\end{equation}}}
\newcommand{\eeq}{\end{equation}}
\newcommand{\nn}{\nonumber}
\newcommand{\bea}{\begin{eqnarray}}
\newcommand{\eea}{\end{eqnarray}}
\newcommand{\bfig}{\begin{figure}}
\newcommand{\efig}{\end{figure}}
\newcommand{\bc}{\begin{center}}
\newcommand{\ec}{\end{center}}

\newcommand{\f}[2]{\frac{#1}{#2}}

\date{}

\begin{document}
\maketitle
\flushbottom

\section{Introduction, physics motivations, and conclusions} \label{1}

In the present paper we work out the interplay between lowest-order perturbative computations in the 't Hooft coupling \cite{tHooft:1973alw}, $g^2=g^2_{YM} N$, operator mixing, renormalization-group (RG) improved ultraviolet (UV) asymptotics of leading-order (LO) nonplanar \footnote{The LO-nonplanar contribution is the next-to-leading one in the large-$N$ expansion.} /planar \footnote{The planar contribution is the leading one in the large-$N$ expansion.} contributions to $2$-point correlators, and nonperturbative large-$N$ 't Hooft expansion of perturbatively massless asymptotically free QCD-like theories (massless QCD-like theories in short) (section \ref{2}), such as pure Yang-Mills (YM) theory, QCD with massless quarks, and their $\mathcal{N}=1$ supersymmetric extensions. \par 
Apart from the intrinsic interest, our main motivation is for the qualitative and quantitative constraints that our UV asymptotic computations set on the -- yet to come -- nonperturbative solution of large-$N$ YM theory and, more generally, of massless QCD-like theories. \par
As concrete examples, we compute, to the lowest perturbative order \footnote{We suppose that all the operators are canonically normalized, in such a way that the perturbative contributions to the $2$-point correlators start from the order $g^0$.} in $g$, the ratios, $r_i$, of LO-nonplanar to planar contributions to the $2$-point correlators in the coordinate representation of certain local gauge-invariant operators (sections \ref{4} and \ref{5}), which mix among themselves under renormalization, in their orthogonal basis (section \ref{3}) in $SU(N)$ YM theory. \par
Though $r_i$ has an intrinsic meaning from the point of view of the large-$N$ expansion, as it does not depend on the normalization of the operators in the planar theory  \cite{tHooft:1973alw}, its very definition depends on the choice of the operators and, specifically, on the existence of an orthogonal basis (section \ref{3}) to the lowest perturbative order, in such a way that its meaning is apparently limited to the lowest order only.\par
Indeed, while to the lowest order $r_i$ is a pure number (section \ref{2}) in massless QCD-like theories, to higher orders and nonperturbatively, the space-time dependence of the LO-nonplanar and planar contributions does not cancel in general in their ratio (section \ref{30}), thus making the very definition of $r_i$ as a pure number in general meaningless beyond the lowest order. \par
Yet, as opposed to the general case, we demonstrate (sections \ref{3} and \ref{30}) according to \cite{Mixing0,MBR} that -- if $\frac{\gamma_0}{\beta_0}$ has no LO-nonplanar contribution \cite{MBR} -- $r_i$ coincides instead with the corresponding ratio in the large-$N$ expansion of the RG-improved UV asymptotics of $2$-point correlators, provided that a certain operator basis exists (sections \ref{2} and \ref{3}), which to the lowest perturbative order coincides with the orthogonal basis (section \ref{3}), where $\frac{\gamma(g)}{\beta(g)}= \frac{\gamma_0}{\beta_0} \frac{1}{g}$ is diagonal and one-loop exact  \cite{Mixing0} to all orders of perturbation theory, with $\gamma(g)$ the anomalous-dimension matrix, $\beta(g)$ the beta function, and $\gamma_0,\beta_0$ their first coefficients respectively. \par
Indeed, in this case, the space-time dependence of LO-nonplanar and planar contributions to $2$-point correlators exactly cancels in the UV asymptotics of their ratio (section \ref{30}).\par
We refer to the above renormalization scheme for $\frac{\gamma(g)}{\beta(g)}$ as to the canonical nonresonant diagonal renormalization scheme introduced in \cite{Mixing0} (section \ref{3}). \par
Its existence depends on the nonresonant condition \cite{Mixing0} in eq. \eqref{Nonres}, for the differences of pairs of eigenvalues of $\frac{\gamma_0}{\beta_0}$, that is generically satisfied \cite{Mixing0}.\par
We classify the operators that mix among themselves under renormalization in massless QCD-like theories in two mutually disjoint classes: The class of operators for which the set of the ratios $r_i$ in the orthogonal basis contains at least a real -- but not rational -- number, which we refer to as the real class, and the class of operators for which all the aforementioned ratios are rational numbers, which we refer to as the rational class.\par
If $r_i$ coincides with the corresponding LO-nonplanar/planar ratio in the actual UV asymptotics mentioned above, the rationality of $r_i$ may have a nonperturbative meaning in the exact solution of large-$N$ massless QCD-like theories. \par
Indeed, we find an infinite set of operators in large-$N$ YM theory that belong to the rational class: For all of the gauge-invariant gluonic twist-$2$ operators \cite{Twist2}, $r_i= - \frac{1}{N^2}=r_{twist-2}$ universally (section \ref{5}), and $\frac{\gamma_0}{\beta_0}$ has no nonplanar contribution (section \ref{5}).\par
Moreover, we verify by means of Mathematica (section \ref{5}) that, for the first $10^3$ twist-$2$ operators ordered by increasing spin, the condition (section \ref{3}) for the existence of the canonical nonresonant diagonal renormalization scheme is satisfied.\par
Therefore, for the twist-$2$ operators above in large-$N$ YM theory, $r_i$ has a nonperturbative meaning.
Indeed, according to fundamental principles \cite{Bochicchio:2016toi} of the large-$N$ expansion \cite{tHooft:1973alw} of YM theory, we point out that such $r_i$ must coincide with the UV asymptotics of the ratio between the glueball self-energy loop \cite{Bochicchio:2016toi} and the glueball tree contribution \cite{Bochicchio:2016toi} to the aforementioned $2$-point correlators in the canonical nonresonant diagonal renormalization scheme.\par
Hence, our main conclusion is that the universality of $r_{twist-2}$ is related to the universality of the effective coupling in the nonperturbative large-$N$ expansion of the YM theory, and that our computations provide a quantitative measure of the strength of such coupling for the twist-$2$ operators in the coordinate representation. \par

\section{Main results and plan of the paper} \label{2}

We now describe the main results in the present paper concerning the technical aspects of our computations. \par 
To the leading and next-to-leading order in perturbation theory, which we refer to as the lowest orders, massless QCD-like theories are actually conformal invariant \cite{ConfQCD}, since the beta function:
\bea
\beta(g)= - \beta_0 g^3 - \beta_1 g^5 + \cdots
\eea
only affects the solution of the Callan-Symanzik equation starting from the order of $g^4$. Therefore, our lowest-order computations are controlled by the conformal symmetry \cite{ConfQCD} to the lowest orders.\par
Specifically, $r_i$ is a pure number because, to the lowest order in $g$, the theory is exactly conformal, and therefore the space-time dependence of the $2$-point correlators is completely fixed by the conformal symmetry and cancels exactly in the ratio between LO-nonplanar and planar contributions. \par
In the concrete examples that we work out, and in general as a rule of thumb, operators that do not mix to order $g^0$, i.e., that are mutually orthogonal to the lowest order in their natural defining basis, turn out to belong to the rational class.\par
On the contrary, operators that do mix to order $g^0$, i.e., that are not mutually orthogonal to the lowest order in their natural defining basis, turn out to belong in general to the real class.
In the present paper we work out concrete examples of both classes.\par
Our first main result is that, by explicit computation, in $SU(N)$ YM theory certain gauge-invariant gluonic dimension-$8$ scalar operators \cite{G1,G2} belong to the real class (section \ref{4}),
while all the gauge-invariant gluonic twist-$2$ operators belong to the rational class (section \ref{5}). \par
Moreover, for the dimension-$8$ scalar operators, $\frac{\gamma_0}{\beta_0}$ has a nonvanishing LO-nonplanar contribution (section \ref{4}), in such a way that the corresponding $r_i$ is not well defined (section \ref{30}) beyond the lowest perturbative order.\par
Incidentally, we also demonstrate (section \ref{4}), as an example of intrinsic interest, that for the aforementioned dimension-$8$ operators the resonant condition in eq. \eqref{Res} holds in the planar theory, but it is lifted to the leading nonplanar order. \par
On the contrary, for all of the gluonic twist-$2$ operators, $r_i$ has the universal rational value (section \ref{5}):
\bea \label{U}
 r_{twist-2}= - \frac{1}{N^2}
 \eea
and $\frac{\gamma_0}{\beta_0}$ has no nonplanar contribution (section \ref{5}). Moreover, the canonical nonresonant diagonal renormalization scheme exists generically \cite{Mixing0}, and, specifically, we verify (section \ref{5}) that it exists for the first $10^3$ twist-$2$ operators. 
Therefore, for such operators, $r_i$ makes sense beyond perturbation theory (section \ref{30}).\par
The computation of $r_i$ involves the construction of the orthogonal basis where $G_0$, the coefficient matrix (section \ref{3}) that sets the normalization of 2-point correlators to order $g^0$, is diagonal. \par
Relatedly, our second main result consists in verifying (sections \ref{4} and \ref{5}) in passing the relation \cite{Mixing2} (section \ref{3}):
\bea \label{1.9}
\gamma_0 G_0 = G_0\gamma_0^T
\eea 
where $\gamma_0^T$ is the transposed matrix of $\gamma_0$,
both for the gluonic dimension-$8$ scalar operators and the gluonic twist-$2$ operators in YM theory. 
By assuming that $\gamma_0$ is diagonalizable (section \ref{3}), eq. \eqref{1.9} allows us the simultaneous diagonalization of $G_0$ \footnote{For the operators that are considered in the present paper, $G_0$ is real symmetric and thus diagonalizable by a change of the operator basis (section \ref{3}).} and $\gamma_0$ (section \ref{3}).\par
In fact, to the best of our knowledge, eq. \eqref{1.9} is verified for the first time in the present paper, as in the defining basis of the dimension-$8$ scalar operators both $\gamma_0$ and $G_0$ are nondiagonal. For this aim we employ the computation of $G_0$ in the present paper (section \ref{4}) and of $\gamma_0$ in \cite{G2} (section \ref{4}).\par
For the twist-$2$ operators eq. \eqref{1.9} is trivially verified, as both $\gamma_0$ and $G_0$ are already diagonal in the defining basis (section \ref{5}).\par
Besides, we work out (section \ref{3}) the close interplay between our lowest-order computation and the actual nonperturbative asymptotics of $2$-point correlators in the case of operator mixing according to \cite{Mixing0,Mixing,Mixing2}.\par
Our third main result consists in demonstrating (section \ref{30}) for Hermitian operators that, if $\frac{\gamma_0}{\beta_0}$ has no LO-nonplanar contribution, the perturbative lowest-order ratio, $r_i$, equals the corresponding ratio in the large-$N$ expansion of the UV asymptotics of $2$-point correlators in a certain canonical nonresonant diagonal renormalization scheme (section \ref{3}), where $\frac{\gamma(g)}{\beta(g)}$ is one-loop exact and diagonal. In such a scheme, $G_0$ and $\gamma_0$ are simultaneously diagonal as a consequence of eq. \eqref{1.9}. \par Moreover, such a scheme  exists generically, provided that the nonresonant condition \cite{Mixing0} for the eigenvalues of $\frac{\gamma_0}{\beta_0}$ in eq. \eqref{Nonres} is satisfied (section \ref{3}). We verify numerically that, for the first $10^3$ twist-$2$ operators, the aforementioned nonresonant condition is actually satisfied (section \ref{5}). \par

\section{A detour through operator mixing} \label{3}

We summarize briefly the results about operator mixing in \cite{Mixing0,Mixing,Mixing2}, both to the lowest orders and to all orders, that are relevant for the present paper. \par
In a massless QCD-like theory, we consider 2-point correlators of renormalized bosonic Hermitian local gauge-invariant operators in Minkowskian space-time, either $PT$ even or odd, with maximal projection of the integer spin $s$ along the direction $p_+= \frac{1}{\sqrt2} (p_0+p_3)$, naive dimension $D=s+\tau$ and twist $\tau$ \cite{gluon}.\par
Moreover, we perform the Wick rotation \cite{gluon} of the aforementioned operators and correlators to Euclidean space-time:
\bea
G^{(s)}_{ik}(x) = \langle O^{(s)}_i(x) O^{(s)}_k(0) \rangle 
\eea
where $O^{(s)}_i(x)$ are the renormalized Euclidean operators with spin projection $s$ along the direction $p_z= \frac{1}{\sqrt2} (p_4+i p_3)$:
\bea
O^{(s)}_i= Z_{ik} O^{(s)}_{Bk}
\eea
with $O^{(s)}_{Bk}$ the bare operators that mix under renormalization \footnote{In fact, gauge-invariant operators also mix with BRST-exact operators and with operators that vanish by the equations of motion (EQM). But correlators of gauge-invariant operators with BRST-exact operators vanish, while correlators with EQM operators reduce to contact terms. Hence, for our purposes it suffices to take into account the mixing of gauge-invariant operators only.}.
In matrix notation:
\bea
O^{(s)}= Z O^{(s)}_B
\eea
where $O^{(s)}$ denotes a column vector whose entries are the operators $O^{(s)}_{i}$.
The Euclidean $2$-point correlators satisfy the Callan-Symanzik equation \cite{C,S} in matrix notation \cite{Mixing0,Mixing2}:
\be
\left(x \cdot \frac{\partial}{\partial x}+\beta(g)\frac{\partial}{\partial g} \right) G^{(s)}+ \gamma(g) G^{(s)} + G^{(s)} \gamma^T(g) =0
\ee 
Its asymptotic solution \cite{Mixing0, Mixing2} reads in Euclidean space-time:
\bea \label{1.8}
G^{(s)}(x) \sim \frac{x_z^{2s}}{(x^2)^{2s+\tau}}  
Z^{(s)}(\sqrt {x^2}, \mu)\mathcal{G}^{(s)}(g(\sqrt {x^2}))Z^{(s)T}(\sqrt {x^2}, \mu)
\eea
and, by analytic continuation \cite{gluon}, $x_z \rightarrow i x_+$, $x^2 \rightarrow - |x|^2 + i \epsilon$, in Minkowskian space-time:
\bea \label{1.80}
(-1)^s \frac{x_+^{2s}}{(-|x|^2+i \epsilon)^{2s+\tau}}  
Z^{(s)}(\sqrt {-x^2+i \epsilon}, \mu)\mathcal{G}^{(s)}(g(\sqrt {-x^2+i \epsilon}))Z^{(s)T}(\sqrt {-x^2+i \epsilon}, \mu)\nonumber \\
\eea
where $x^2$ and $|x|^2$ are the Euclidean and Minkowskian squared distance \cite{gluon} respectively.
In eqs. \eqref{1.8} and \eqref{1.80} the space-time dependence of the first factor is uniquely fixed by translational invariance, dimensional analysis, the absence of perturbative scales to the lowest order in massless QCD-like theories, and, respectively, the Euclidean and Lorentz symmetry. \par
$Z^T$ is the transposed matrix of $Z$, $\mathcal{G}^{(s)}(g(x))$ is a dimensionless RG-invariant matrix function -- which is symmetric, as the operators are bosonic and either $PT$ even or odd, and real, as the operators are Hermitian -- of the running coupling only \cite{Mixing}, and:
\begin{equation}
\label{01.9}
Z^{(s)}(x, \mu)=P\exp\left(-\int_{g(x)}^{g(\mu)}\frac{\gamma^{(s)}(g)}{\beta(g)}dg\right) \equiv Z^{(s)}(\sqrt {x^2}, \mu)
\end{equation}
where $P$ denotes the path ordering of the exponential and $g(\mu)$ and $g(x)\equiv  g(\sqrt {x^2}) $ are short notations for the running coupling at the corresponding scales, $g(\frac{\mu}{\Lambda_{RG}})$ and $g(x \Lambda_{RG}) \equiv g(\sqrt {x^2}\Lambda_{RG}) $:
\begin{equation}
\label{1.12}
g^2(x \Lambda_{RGI})  \sim \dfrac{1}{\beta_0\log(\frac{1}{x^2 \Lambda_{RGI}^2})} \left(1-\dfrac{\beta_1}{\beta_0^2} \dfrac{\log\log(\frac{1}{x^2 \Lambda_{RGI}^2})}{\log(\frac{1}{x^2 \Lambda_{RGI}^2})}\right)
\end{equation}
with $\Lambda_{RG}$ the RG-invariant scale and $\gamma^{(s)}(g)$ the matrix of the anomalous dimensions:
\begin{equation}
\label{1.6}
\gamma^{(s)}(g)=- \frac{\partial Z^{(s)}}{\partial \log \mu} Z^{(s) -1}=  \gamma^{(s)}_{0} g^2  + \gamma^{(s)}_{1} g^4 +\cdots
\end{equation}
In the following, for brevity, we omit the superscript $^{(s)}$, since the spin $s$ has no further role.
Both $\mathcal{G}(g(x))$ and $Z(x,\mu)$ admit the perturbative expansion in terms of the renormalized coupling $g=g(\mu)$:
\bea \label{a}
\mathcal{G}(g) = G_0 + g^2 G_1 + \cdots
\eea
with $G_0$ real symmetric, and:
\bea
Z(x, \mu)= I - g^2 \gamma_0\log\sqrt {x^2 \mu^2}+ \cdots
\end{eqnarray}
with $I$ the identity matrix.
Moreover, in a massless QCD-like theory, it holds \cite{Mixing2}:
\beq
\label{2.35}
\gamma_0 G_0 = G_0\gamma_0^T
\eeq
that is a consequence \cite{Mixing2} of the existence of a scalar product induced by the conformal structure (section \ref{2}) to the lowest orders. \par
We briefly summarize the proof \cite{Mixing2} of eq. \eqref{2.35} for Hermitian scalar gauge-invariant operators.
In a conformal field theory, $G_{conf}(x)$ satisfies the Callan-Symanzik equation:
\beq
\label{CSconf}
x\cdot\dfrac{\partial}{\partial x}G_{conf}(x) + \Delta G_{conf}(x) + G_{conf}(x)\Delta^T = 0
\eeq
with $\Delta$ the matrix of the conformal dimensions,
whose general solution for scalar operators is:
\beq
\label{G2conf}
G_{conf}(x) = \langle O(x)O(0) \rangle = e^{-\Delta \log \sqrt {x^2 \mu^2}} \mathcal{G} e^{-\Delta^T \log \sqrt {x^2 \mu^2}}
\eeq
in matrix notation, where $\mathcal{G}$ is a matrix independent of space-time and the tensor product between repeated $O$ is understood. If $\Delta$ is diagonalizable, the dependence on the scale $\mu$ may be reabsorbed into a redefinition of $\mathcal{G}$. Otherwise, if $\Delta$ is nondiagonalizable, a logarithmic \cite{CFT2} conformal field theory arises. \par
Moreover, in a conformal field theory the operators/states correspondence holds \cite{CFT1,CFT2}: 
\begin{eqnarray}
O(0)\vert 0 \rangle &=& \vert O_{in}\rangle \nonumber \\
\langle O_{out}\vert &=& \lim_{x\rightarrow \infty}\langle 0\vert e^{2\Delta\log \sqrt {x^2 \mu^2}} O(x)
\end{eqnarray}
As a consequence, the scalar product in matrix notation reads:
\begin{eqnarray}
\label{ScalarP}
\langle O_{out}\vert O_{in} \rangle &=& \lim_{x\rightarrow \infty} \langle 0 \vert e^{2\Delta\log \sqrt {x^2 \mu^2}} O(x)O(0)\vert 0 \rangle \nonumber \\
& = &\lim_{x\rightarrow \infty}  e^{2\Delta\log \sqrt {x^2 \mu^2}} e^{-\Delta \log \sqrt {x^2 \mu^2}} \mathcal{G} e^{-\Delta^T \log \sqrt {x^2 \mu^2}}  \nonumber \\
&=& \lim_{x\rightarrow \infty}  e^{\Delta\log \sqrt {x^2 \mu^2}} \mathcal{G} e^{-\Delta^T\log \sqrt {x^2 \mu^2}} 
\end{eqnarray}
In order to be well defined, the scalar product in eq. \eqref{ScalarP} must be independent of the variable $\sqrt {x^2 \mu^2}$.
To the lowest orders in massless QCD-like theories we get:
\begin{eqnarray}
\Delta(g) & = &  D \, I + g^2 \gamma_0 + \cdots \\ \nonumber
\mathcal{G}(g) & = & G_0 + g^2 G_1 + \cdots
\end{eqnarray}
in the conformal renormalization scheme \cite{ConfQCD}. Expanding the scalar product above to the lowest orders, we obtain:
\begin{eqnarray}
\label{2.34}
\langle O_{out} \vert O_{in} \rangle &&= \left(I +g^2\gamma_0\log \sqrt {x^2 \mu^2}\right)\mathcal{G}(g)\left(I - g^2\gamma_0^T\log \sqrt {x^2 \mu^2}\right) \nonumber \\
&&=  (G_0 + g^2G_1) +g^2\left(\gamma_0 G_0 - G_0\gamma_0^T\right)\log \sqrt {x^2 \mu^2} +\cdots
\end{eqnarray}
The independence from $\sqrt {x^2 \mu^2}$ then implies eq. \eqref{2.35}. Interestingly, eq. \eqref{2.35} mixes \cite{Mixing2} the lowest orders of perturbation theory.\par
For the change of basis:
\bea \label{b}
O'(x)=S  O(x)
\eea
with $S$ a constant invertible matrix, $G(x)$ and $Z(x, \mu)$ transform \cite{Mixing2} as:
\be \label{G}
G'(x) = S G(x) S^T
\ee
and:
\bea
Z'(x, \mu)=SZ(x, \mu)S^{-1}
\eea
Correspondingly, $G_0$ and $\gamma_0$ transform as:
\bea \label{trr}
G'_0= S G_0 S^T
\eea
and:
\be \label{gamma}
\gamma'_0= S \gamma_0 S^{-1}
\ee 
Consistently, eq. \eqref{2.35} is covariant for the aforementioned change of basis. \par
If $\gamma_{0}$ is diagonalizable by a change of basis, as it should be \cite{Mixing,Mixing2} for Hermitian operators in the gauge-invariant sector of a massless QCD-like theory, unless a nonunitary \cite{CFT2} conformal field theory arises \cite{Mixing,Mixing2} to the lowest orders, by eq. \eqref{2.35} $G_0$ commutes with $\gamma_0$ in the diagonal basis.
Moreover, as $G_0$ is a real symmetric matrix, it is diagonalizable \cite{Mixing2} by the transformation in eq. \eqref{trr}.
Therefore, under the above assumption, a basis exists where $\gamma_0$ and $G_0$ are both diagonal. \par 
Thus, an orthogonal basis of operators with well defined conformal dimensions, $\Delta_i= D +\gamma_{0i} g^2$, exists to the lowest orders. 
These lowest-orders features can be combined with the general solution in eq. \eqref{1.8} to get asymptotic estimates for the $2$-point correlators.\par
For this purpose, it is needed an all-orders extension of the lowest-orders diagonal basis for $\gamma_0$ and $G_0$, i.e., the existence of a canonical diagonal renormalization scheme where:
\begin{equation}
\label{1.70}
-\frac{\gamma_{ik}(g)}{\beta(g)} = \sum_{s} \frac{\partial Z_{is}}{\partial g} (Z^{-1})_{sk}=   \frac{ \gamma_{0ik}}{\beta_0 g}   + \cdots
\end{equation}
is diagonal and one-loop exact \cite{Mixing0}:
\begin{equation}
\label{1.7}
\frac{ \gamma_{0i}}{\beta_0 g} = \frac{\partial Z_{i}}{\partial g} Z_{i}^{-1}
\end{equation}
with $\gamma_{0i}$ and $Z_i$ the eigenvalues of the corresponding matrices. \par
We may wonder whether such a scheme actually exists. The general answer to this question has been worked out in \cite{Mixing0,Mixing} on the basis of the following simple, but fundamental, observations in \cite{Mixing0}.\par
It has been pointed out in \cite{Mixing0} that renormalization can be interpreted in a differential geometric setting, where a (finite) change of renormalization scheme, i.e., a coupling-dependent change of the operator basis:
\bea \label{b}
O'(x)=S(g) O(x)
\eea
is interpreted as a matrix-valued (formal) holomorphic invertible gauge transformation $S(g)$.
Accordingly, the matrix valued ratio:
\bea
-\frac{\gamma(g)}{\beta(g)}= \frac{\gamma_0}{\beta_0} \frac{1}{g}+\cdots
\eea
that occurs in the system of ordinary differential equations defining $Z(x, \mu)$:
\bea \label{1.700}
\big(\frac{\partial}{\partial g} +\frac{\gamma(g)}{\beta(g)}\big) Z(x, \mu) =0
\eea
is interpreted \cite{Mixing0} as a (formal) meromorphic connection $A(g)$, with a simple pole at $g=0$ -- i.e., a Fuchsian singularity --, that for the gauge transformation in eq. \eqref{b} transforms as:
\bea
A'(g)= S(g)A(g)S^{-1}(g)+ \frac{\partial S(g)}{\partial g} S^{-1}(g)
\eea
Morevover,
\bea
\mathcal{D}= \frac{\partial}{\partial g} +\frac{\gamma(g)}{\beta(g)} =  \frac{\partial}{\partial g} -A(g)
\eea
is interpreted as the corresponding covariant derivative. As a consequence, $Z(x, \mu)$ is 
interpreted \cite{Mixing0} as a Wilson line associated to the aforementioned connection:
\bea
Z(x, \mu)=P\exp\left(-\int^{g(\mu)}_{g(x)}\frac{\gamma(g)}{\beta(g)}dg\right)=P\exp\left(\int ^{g(\mu)}_{g(x)} A(g) \, dg\right)
\eea
that transforms as:
\bea
Z'(x, \mu)= S(g(\mu)) Z(x, \mu) S^{-1}(g(x))
\eea
for the gauge transformation $S(g)$.\par
The above observations in \cite{Mixing0}, though likely well known, have opened the way to apply in the framework of operator mixing the theory of canonical forms \cite{Mixing0,Mixing} for systems of differential equations with Fuchsian singularities obtained by means of (formal) holomorphic gauge transformations and, specifically, the Poincar\'e -Dulac theorem  \cite{Mixing0,Mixing}.\par
Indeed, the easiest way \cite{Mixing0,Mixing} to compute the UV asymptotics of $Z(x, \mu)$ consists in setting the connection, $A(g)=-\frac{\gamma(g)}{\beta(g)}$, in canonical form by a suitable gauge transformation.\par
On the basis of the Poincar\'e-Dulac theorem applied to the system of differential equations for $Z(x, \mu)$ in eq. \eqref{1.700}, it turns out \cite{Mixing0} that, if the following nonresonant condition for the eigenvalues of $\frac{\gamma_0}{\beta_0}$ is satisfied, a (formal) holomorphic gauge transformation exists that sets eq. \eqref{1.700} in the diagonal canonical form in eq. \eqref{1.7}. \par
The sufficient \cite{Mixing0} condition for the mixing associated to eq. \eqref{1.700} to be nonresonant is that any two eigenvalues $\lambda_1, \lambda_2, \cdots $ of the matrix $\frac{\gamma_0}{\beta_0}$, in nonincreasing order $\lambda_1 \geq \lambda_2 \geq \cdots$, do not differ by a positive even integer, i.e.:
\bea \label{Nonres}
\lambda_i -\lambda_j -2k  \neq 0
\eea
for $i\leq j$ and $k$ a positive integer. \par
Otherwise, in the resonant case, $Z(x, \mu)$ is not diagonalizable \cite{Mixing0} by a holomorphic gauge transformation, and a more complicated canonical form \cite{Mixing} for $-\frac{\gamma(g)}{\beta(g)}$ and UV asymptotics for $Z(x, \mu)$ arises \cite{Mixing} from the Poincar\'e-Dulac  theorem. 
The necessary condition for the mixing to be resonant is that:
\bea \label{Res}
\lambda_i -\lambda_j -2k  = 0
\eea
for some $i < j$ and $k$ a positive integer. \par
Thus, the nonresonant mixing -- for which the diagonal canonical renormalization scheme exists -- is the generic case \cite{Mixing0}.
In the nonresonant diagonal scheme where eq. \eqref{1.7} holds, $Z(x, \mu)$ is diagonal and its eigenvalues, $Z_{i}(x, \mu)$, are computed by:
\begin{equation}
\label{01.9}
Z_{i}(x, \mu)=\exp\left(\int^{g(\mu)}_{g(x)}\frac{\gamma_{0i}}{\beta_0 g}dg\right) = \left(\frac{g(\mu)}{g(x)}\right)^{\frac{\gamma_{0i}}{\beta_0}}
\end{equation}
Correspondingly, eq. \eqref{1.8} reads:
\bea \label{10}
G_{ii}(x) &\sim &  \frac{x_z^{2s}}{(x^2)^{2s+\tau}}  
Z_{i}(x, \mu) \mathcal{G}_{ii}(g(x)) Z_{i}^T(x, \mu)
\eea
where no sum on the index $i$ is understood. Eqs. \eqref{10}, \eqref{01.9} and \eqref{a} lead to the asymptotic estimates:
\bea \label{1.11}
G_{ii}(x) &\sim&    \frac{x_z^{2s}}{(x^2)^{2s+\tau}}  
 \left(\frac{g(\mu)}{g(x)}\right)^{\frac{2\gamma_{0i}}{\beta_0}} G_{0i}
\eea
Thus, provided that the canonical nonresonant diagonal renormalization scheme exists, we have reduced \cite{Mixing0} the asymptotics of operator mixing to the multiplicatively renormalizable case.

\section{UV asymptotics of the nonplanar/planar contributions to $2$-point correlators}  \label{30}

Now, assuming that the canonical nonresonant diagonal renormalization scheme exists, we expand in powers of $\frac{1}{N}$ the rhs of eq. \eqref{1.11}. 
According to \cite{MBR}, the nonplanar contributions, $[\frac{\gamma_{0i}}{\beta_0}]^{NP}$, in:
 \bea
 \frac{\gamma_{0i}}{\beta_0}= [\frac{\gamma_{0i}}{\beta_0}]^P+ [\frac{\gamma_{0i}}{\beta_0}]^{NP}
 \eea
produce terms involving logs of the running coupling:
\bea \label{1.12}
G_{ii}(x) &\sim&  \frac{x_z^{2s}}{(x^2)^{2s+\tau}}     \left(\frac{g(\mu)}{g(x)}\right)^{2 [\frac{\gamma_{0i}}{\beta_0}]^P+2 [\frac{\gamma_{0i}}{\beta_0}]^{NP}} (G^P_{0i} + G^{NP}_{0i}) \nonumber \\
 &\sim&    \frac{x_z^{2s}}{(x^2)^{2s+\tau}}   \left(\frac{g(\mu)}{g(x)}\right)^{2 [\frac{\gamma_{0i}}{\beta_0}]^P} \left(1+ 2 [\frac{\gamma_{0i}}{\beta_0}]^{NP} \log\left(\frac{g(\mu)}{g(x)}\right) + \cdots \right)  (G^P_{0i} + G^{NP}_{0i}) \nonumber \\
 &\sim&    \frac{x_z^{2s}}{(x^2)^{2s+\tau}}    \left(\frac{g(\mu)}{g(x)}\right)^{2 [\frac{\gamma_{0i}}{\beta_0}]^P}  G^P_{0i} \nonumber \\
  &+&         \frac{x_z^{2s}}{(x^2)^{2s+\tau}}    \left(\frac{g(\mu)}{g(x)}\right)^{2 [\frac{\gamma_{0i}}{\beta_0}]^P}  \left(G^{NP}_{0i} + G^P_{0i} \left( 2 [\frac{\gamma_{0i}}{\beta_0}]^{NP}  \log\left(\frac{g(\mu)}{g(x)} \right)+\cdots  \right) + \cdots\right) \nonumber \\
 \eea
with:
\bea
G_{0i}=G^P_{0i}+G^{NP}_{0i}
\eea
Hence:
\bea
\frac{G^{NP}_{ii}(x)}{G^{P}_{ii}(x)} \sim \frac{G^{NP}_{0i}}{G^{P}_{0i}} + \left( 2 [\frac{\gamma_{0ii}}{\beta_0}]^{NP}  \log\left(\frac{g(\mu)}{g(x)} \right)+\cdots  \right)+\cdots
\eea
As a consequence:
\bea
 r_i = \frac{G^{LO-NP}_{0i}}{G^{P}_{0i}}\sim \frac{G^{LO-NP}_{ii}(x)}{G^{P}_{ii}(x)} 
\eea
provided that $G^{LO-NP}_{0i}$ and  $G^{P}_{0i}$ do not vanish, and $[\frac{\gamma_{0i}}{\beta_0}]^{LO-NP}=0$.

\section{Gluonic dimension-$8$ scalar operators} \label{4}

\subsection{The defining basis}

We consider in $SU(N)$ YM theory the set -- defined in \cite{G2} -- of bare dimension-$8$ gauge-invariant Hermitian scalar operators:
\bea
\label{dim8set}
\mathcal{O}_{B841} & = & \f{1}{N^4}F^a_{\mu\sigma}F^{a \, \mu\rho}F^{b \, \sigma\nu}F^b_{\rho\nu} \;\;\;  \;\;\; \mathcal{O}_{B842} =  \f{1}{N^4}F^a_{\mu\sigma}F^{b \, \mu\rho}F^{b \, \sigma \nu}F^a_{\rho\nu} \nn \\
\mathcal{O}_{B843} & = &  \f{1}{N^4}F^a_{\mu\sigma}F^a_{\nu\rho}F^{b \, \sigma\mu}F^{b \, \rho\nu} \;\;\;  \;\;\; \mathcal{O}_{B844} =  \f{1}{N^4}F^a_{\mu\sigma}F^b_{\nu\rho}F^{a \, \sigma\mu}F^{b \, \rho\nu} \nn \\
\mathcal{O}_{B845} & = &  \f{1}{N^4}d_4^{abcd}F^a_{\mu\sigma}F^{b \, \mu\sigma}F^c_{\nu\sigma}F^{d \, \nu\rho} \;\;\; \;\;\; \mathcal{O}_{B846} =  \f{1}{N^4}d_4^{abcd}F^a_{\mu\sigma}F^{c \, \mu\rho}F^{b \, \nu\sigma}F^d_{\nu\rho} \nn \\
\mathcal{O}_{B847} & = &  \f{1}{N^4}d_4^{acbd}F^a_{\mu\sigma}F^{b \, \mu\sigma}F^c_{\nu\rho}F^{d \, \nu\rho} \;\;\; \;\;\; \mathcal{O}_{B848} =  \f{1}{N^4}d_4^{abdc}F^a_{\mu\sigma}F^{c \, \mu\rho}F^{b \, \nu\sigma}F^d_{\nu\rho}
\eea
where $F_{\mu\nu}^a$ is the field strength:
\be
F_{\mu\nu}^a = \partial_\mu A^a_\nu - \partial_{\nu} A^a_{\mu} - g f^{abc}A^b_{\mu}A^c_\nu
\ee
$d_4^{abcd}$ is the rank-$4$ tensor:
\be
d_4^{abcd} = d^{abe}d^{dce}
\ee   
with $f^{abc}$ and $d^{abc}$ the antisymmetric and symmetric tensors respectively:
\bea
&&\left[T^a, T^b\right] = i f^{abc} T^c \nn \\
&& \left\{T^a,T^b\right\} = \frac{1}{N}\delta^{ab} I + d^{abc}T^c
\eea
associated to the generators, $T^a$, of the Lie algebra of $SU(N)$ in the fundamental representation, normalized as:
\bea
\Tr(T^aT^b)=\frac{1}{2} \delta^{ab}
\eea
We refer to the operators, $\mathcal{O}_{B841}\cdots\mathcal{O}_{B844}$, and $\mathcal{O}_{B845}\cdots \mathcal{O}_{B848}$, as to double-trace and single-trace operators respectively. These operators mix among themselves under renormalization in perturbation theory \cite{G1,G2}:
\be
\mathcal{O} = Z\mathcal{O}_B
\ee 
where $\mathcal{O}$ is the column vector containing the renormalized operators defined in eq. \eqref{dim8set}, whose transposed, $\mathcal{O}^T$, reads:
\be
\mathcal{O}^T = \left(\mathcal{O}_{841} \, \mathcal{O}_{8412} \, \mathcal{O}_{843} \,\mathcal{O}_{844} \,\mathcal{O}_{845} \,\mathcal{O}_{846} \,\mathcal{O}_{847} \,\mathcal{O}_{848}\right)
\ee
with $\mathcal{O}_B$ the vector of the bare operators.

\subsection{$\gamma_0$ in the defining basis}

The matrix $\gamma(g)$ of the anomalous dimensions for the above operators has been computed to the order of $g^2$ in \cite{G2}. From \cite{G2} we extract the large-$N$ expansion of $\gamma_0$:
\be
\label{gamma0}
\gamma_0 = \sum_{k=0}^3\frac{1}{N^k}\gamma_{0k}
\ee
where the matrices $\gamma_{0k}$ read:
\bea
\label{gamma0M}
\gamma_{00} & = & \dfrac{1}{(4\pi)^2} \left(
\begin{array}{cccccccc}
 0 & 0 & 0 & \frac{11}{6} & 0 & 0 & 0 & 0 \\
 \frac{14}{3} & \frac{10}{3} & -4 & -\frac{1}{6} & 0 & 0 & 0 & 0 \\
 \frac{28}{3} & -8 & -\frac{2}{3} & -\frac{1}{3} & 0 & 0 & 0 & 0 \\
 0 & 0 & 0 & \frac{22}{3} & 0 & 0 & 0 & 0 \\
 0 & 0 & 0 & 0 & \frac{5}{2} & 6 & \frac{2}{3} & -\frac{16}{3} \\
 0 & 0 & 0 & 0 & \frac{4}{3} & 1 & -1 & 1 \\
 0 & 0 & 0 & 0 & \frac{25}{12} & \frac{19}{3} & -1 & -\frac{16}{3} \\
 0 & 0 & 0 & 0 & \frac{5}{6} & \frac{8}{3} & -2 & \frac{4}{3} \\
\end{array}
\right) \nn \\ \nn \\
\gamma_{01} & = & \dfrac{1}{(4\pi)^2}\left(
\begin{array}{cccccccc}
 0 & 0 & 0 & 0 & \frac{11}{3} & -\frac{4}{3} & -\frac{11}{3} & \frac{4}{3} \\
 0 & 0 & 0 & 0 & \frac{11}{3} & \frac{2}{3} & -\frac{11}{3} & -\frac{2}{3} \\
 0 & 0 & 0 & 0 & \frac{25}{3} & \frac{34}{3} & -\frac{25}{3} & -\frac{34}{3} \\
 0 & 0 & 0 & 0 & -2 & 28 & 2 & -28 \\
 28 & -28 & 2 & -2 & 0 & 0 & 0 & 0 \\
 -\frac{4}{3} & \frac{4}{3} & -\frac{11}{3} & \frac{11}{3} & 0 & 0 & 0 & 0 \\
 \frac{34}{3} & -\frac{34}{3} & -\frac{25}{3} & \frac{25}{3} & 0 & 0 & 0 & 0 \\
 \frac{2}{3} & -\frac{2}{3} & -\frac{11}{3} & \frac{11}{3} & 0 & 0 & 0 & 0 \\
\end{array}
\right) \nn \\ \nn \\
\gamma_{02} & = & \dfrac{1}{(4\pi)^2}\left(
\begin{array}{cccccccc}
 -\frac{8}{3} & \frac{8}{3} & -\frac{22}{3} & \frac{22}{3} & 0 & 0 & 0 & 0 \\
 \frac{4}{3} & -\frac{4}{3} & -\frac{22}{3} & \frac{22}{3} & 0 & 0 & 0 & 0 \\
 \frac{68}{3} & -\frac{68}{3} & -\frac{50}{3} & \frac{50}{3} & 0 & 0 & 0 & 0 \\
 56 & -56 & 4 & -4 & 0 & 0 & 0 & 0 \\
 0 & 0 & 0 & 0 & 4 & -56 & -4 & 56 \\
 0 & 0 & 0 & 0 & -\frac{22}{3} & \frac{8}{3} & \frac{22}{3} & -\frac{8}{3} \\
 0 & 0 & 0 & 0 & -\frac{50}{3} & -\frac{68}{3} & \frac{50}{3} & \frac{68}{3} \\
 0 & 0 & 0 & 0 & -\frac{22}{3} & -\frac{4}{3} & \frac{22}{3} & \frac{4}{3} \\
\end{array}
\right) \nn \\ \nn \\
\gamma_{03} & = & \dfrac{1}{(4\pi)^2}\left(
\begin{array}{cccccccc}
 0 & 0 & 0 & 0 & 0 & 0 & 0 & 0 \\
 0 & 0 & 0 & 0 & 0 & 0 & 0 & 0 \\
 0 & 0 & 0 & 0 & 0 & 0 & 0 & 0 \\
 0 & 0 & 0 & 0 & 0 & 0 & 0 & 0 \\
 -112 & 112 & -8 & 8 & 0 & 0 & 0 & 0 \\
 \frac{16}{3} & -\frac{16}{3} & \frac{44}{3} & -\frac{44}{3} & 0 & 0 & 0 & 0 \\
 -\frac{136}{3} & \frac{136}{3} & \frac{100}{3} & -\frac{100}{3} & 0 & 0 & 0 & 0 \\
 -\frac{8}{3} & \frac{8}{3} & \frac{44}{3} & -\frac{44}{3} & 0 & 0 & 0 & 0 \\
\end{array}
\right)
\eea
\par
\subsection{$G_0$ in the defining basis}

We compute $G_0$ for the operator basis in eq. \eqref{dim8set} by means of a FORM \cite{Ruijl:2017dtg} implementation of the Wick theorem. The large-$N$ expansion of $G_0$ reads:
\be
\label{A0}
G_0 = \sum_{k=0}^6 \frac{1}{N^k} G_{0k}
\ee
with the matrices $G_{0k}$ given by:
\bea
\label{A0M}
G_{00} & = & \dfrac{1}{\pi^8}\left(
\begin{array}{cccccccc}
 576 & 384 & 768 & 1152 & 0 & 0 & 0 & 0 \\
 384 & 768 & 1152 & 384 & 0 & 0 & 0 & 0 \\
 768 & 1152 & 2688 & 768 & 0 & 0 & 0 & 0 \\
 1152 & 384 & 768 & 4608 & 0 & 0 & 0 & 0 \\
 0 & 0 & 0 & 0 & 5376 & 1920 & 3456 & 1536 \\
 0 & 0 & 0 & 0 & 1920 & 1056 & 1728 & 960 \\
 0 & 0 & 0 & 0 & 3456 & 1728 & 4416 & 1920 \\
 0 & 0 & 0 & 0 & 1536 & 960 & 1920 & 1152 \\
\end{array}
\right) \nn \\ \nn \\
G_{01} & = & \dfrac{1}{\pi^8}\left(
\begin{array}{cccccccc}
 0 & 0 & 0 & 0 & 1536 & 960 & 1920 & 1152 \\
 0 & 0 & 0 & 0 & 2304 & 1152 & 1536 & 768 \\
 0 & 0 & 0 & 0 & 5376 & 1920 & 3456 & 1536 \\
 0 & 0 & 0 & 0 & 1536 & 1536 & 5376 & 2304 \\
 1536 & 2304 & 5376 & 1536 & 0 & 0 & 0 & 0 \\
 960 & 1152 & 1920 & 1536 & 0 & 0 & 0 & 0 \\
 1920 & 1536 & 3456 & 5376 & 0 & 0 & 0 & 0 \\
 1152 & 768 & 1536 & 2304 & 0 & 0 & 0 & 0 \\
\end{array}
\right) \nn \\ \nn \\
G_{02} & = & \dfrac{1}{\pi^8}\left(
\begin{array}{cccccccc}
 -192 & 384 & 384 & -768 & 0 & 0 & 0 & 0 \\
 384 & -768 & -768 & 1536 & 0 & 0 & 0 & 0 \\
 384 & -768 & -1920 & 3840 & 0 & 0 & 0 & 0 \\
 -768 & 1536 & 3840 & -7680 & 0 & 0 & 0 & 0 \\
 0 & 0 & 0 & 0 & -54528 & -23424 & -52608 & -23040 \\
 0 & 0 & 0 & 0 & -23424 & -13344 & -23232 & -13248 \\
 0 & 0 & 0 & 0 & -52608 & -23232 & -53568 & -23424 \\
 0 & 0 & 0 & 0 & -23040 & -13248 & -23424 & -13440 \\
\end{array}
\right) \nn \\ \nn \\
G_{03} & = & \dfrac{1}{\pi^8}\left(
\begin{array}{cccccccc}
 0 & 0 & 0 & 0 & -7680 & -4800 & -9600 & -5760 \\
 0 & 0 & 0 & 0 & -11520 & -5760 & -7680 & -3840 \\
 0 & 0 & 0 & 0 & -26880 & -9600 & -17280 & -7680 \\
 0 & 0 & 0 & 0 & -7680 & -7680 & -26880 & -11520 \\
 -7680 & -11520 & -26880 & -7680 & 0 & 0 & 0 & 0 \\
 -4800 & -5760 & -9600 & -7680 & 0 & 0 & 0 & 0 \\
 -9600 & -7680 & -17280 & -26880 & 0 & 0 & 0 & 0 \\
 -5760 & -3840 & -7680 & -11520 & 0 & 0 & 0 & 0 \\
\end{array}
\right) \nn \\ \nn \\
G_{04} & = & \dfrac{1}{\pi^8}\left(
\begin{array}{cccccccc}
 -384 & -768 & -1152 & -384 & 0 & 0 & 0 & 0 \\
 -768 & 0 & -384 & -1920 & 0 & 0 & 0 & 0 \\
 -1152 & -384 & -768 & -4608 & 0 & 0 & 0 & 0 \\
 -384 & -1920 & -4608 & 3072 & 0 & 0 & 0 & 0 \\
 0 & 0 & 0 & 0 & 159744 & 76800 & 190464 & 82944 \\
 0 & 0 & 0 & 0 & 76800 & 44544 & 79872 & 46080 \\
 0 & 0 & 0 & 0 & 190464 & 79872 & 175104 & 76800 \\
 0 & 0 & 0 & 0 & 82944 & 46080 & 76800 & 43008 \\
\end{array}
\right) \nn \\ \nn \\
G_{05} & = & \dfrac{1}{\pi^8}\left(
\begin{array}{cccccccc}
 0 & 0 & 0 & 0 & 6144 & 3840 & 7680 & 4608 \\
 0 & 0 & 0 & 0 & 9216 & 4608 & 6144 & 3072 \\
 0 & 0 & 0 & 0 & 21504 & 7680 & 13824 & 6144 \\
 0 & 0 & 0 & 0 & 6144 & 6144 & 21504 & 9216 \\
 6144 & 9216 & 21504 & 6144 & 0 & 0 & 0 & 0 \\
 3840 & 4608 & 7680 & 6144 & 0 & 0 & 0 & 0 \\
 7680 & 6144 & 13824 & 21504 & 0 & 0 & 0 & 0 \\
 4608 & 3072 & 6144 & 9216 & 0 & 0 & 0 & 0 \\
\end{array}
\right) \nn \\ \nn \\
G_{06} & = & \dfrac{1}{\pi^8}\left(
\begin{array}{cccccccc}
 0 & 0 & 0 & 0 & 0 & 0 & 0 & 0 \\
 0 & 0 & 0 & 0 & 0 & 0 & 0 & 0 \\
 0 & 0 & 0 & 0 & 0 & 0 & 0 & 0 \\
 0 & 0 & 0 & 0 & 0 & 0 & 0 & 0 \\
 0 & 0 & 0 & 0 & -110592 & -55296 & -141312 & -61440 \\
 0 & 0 & 0 & 0 & -55296 & -32256 & -58368 & -33792 \\
 0 & 0 & 0 & 0 & -141312 & -58368 & -125952 & -55296 \\
 0 & 0 & 0 & 0 & -61440 & -33792 & -55296 & -30720 \\
\end{array}
\right)
\eea

\subsection{Relation between $\gamma_0$ and $G_0$ in the defining basis}

We verify eq. \eqref{2.35} to all orders of $\frac{1}{N}$ by employing eqs. \eqref{gamma0}, \eqref{gamma0M}, \eqref{A0} and \eqref{A0M} for $\gamma_0$ and $G_0$ respectively.
Incidentally, this is a powerful check both of the computation of $G_0$ in the present paper and of $\gamma_0$ in \cite{G2}. \par
Yet, for reasons of space, we only display the verification of eq. \eqref{2.35} in the planar theory. The planar contributions to $\gamma_0$ and $G_0$ are given by:
\bea
\gamma_{00} & = & \dfrac{1}{(4\pi)^2}\left(
\begin{array}{cccccccc}
 0 & 0 & 0 & \frac{11}{6} & 0 & 0 & 0 & 0 \\
 \frac{14}{3} & \frac{10}{3} & -4 & -\frac{1}{6} & 0 & 0 & 0 & 0 \\
 \frac{28}{3} & -8 & -\frac{2}{3} & -\frac{1}{3} & 0 & 0 & 0 & 0 \\
 0 & 0 & 0 & \frac{22}{3} & 0 & 0 & 0 & 0 \\
 0 & 0 & 0 & 0 & \frac{5}{2} & 6 & \frac{2}{3} & -\frac{16}{3} \\
 0 & 0 & 0 & 0 & \frac{4}{3} & 1 & -1 & 1 \\
 0 & 0 & 0 & 0 & \frac{25}{12} & \frac{19}{3} & -1 & -\frac{16}{3} \\
 0 & 0 & 0 & 0 & \frac{5}{6} & \frac{8}{3} & -2 & \frac{4}{3} \\
\end{array}
\right)
\eea
and:
\bea
G_{00} & = & \dfrac{1}{\pi^8}\left(
\begin{array}{cccccccc}
 576 & 384 & 768 & 1152 & 0 & 0 & 0 & 0 \\
 384 & 768 & 1152 & 384 & 0 & 0 & 0 & 0 \\
 768 & 1152 & 2688 & 768 & 0 & 0 & 0 & 0 \\
 1152 & 384 & 768 & 4608 & 0 & 0 & 0 & 0 \\
 0 & 0 & 0 & 0 & 5376 & 1920 & 3456 & 1536 \\
 0 & 0 & 0 & 0 & 1920 & 1056 & 1728 & 960 \\
 0 & 0 & 0 & 0 & 3456 & 1728 & 4416 & 1920 \\
 0 & 0 & 0 & 0 & 1536 & 960 & 1920 & 1152 \\
\end{array}
\right)
\eea
respectively. We evaluate both sides of eq. \eqref{2.35} in the planar theory, thus verifying that they are actually equal:
\be
\gamma_{00}  G_{00} = G_{00} \gamma^T_{00}=\dfrac{1}{\pi^8}\dfrac{1}{(4\pi)^2}\left(
\begin{array}{cccccccc}
 2112 & 704 & 1408 & 8448 & 0 & 0 & 0 & 0 \\
 704 & -320 & -3456 & 2816 & 0 & 0 & 0 & 0 \\
 1408 & -3456 & -4096 & 5632 & 0 & 0 & 0 & 0 \\
 8448 & 2816 & 5632 & 33792 & 0 & 0 & 0 & 0 \\
 0 & 0 & 0 & 0 & 19072 & 7168 & 11712 & 4736 \\
 0 & 0 & 0 & 0 & 7168 & 2848 & 3840 & 2240 \\
 0 & 0 & 0 & 0 & 11712 & 3840 & 3488 & 1216 \\
 0 & 0 & 0 & 0 & 4736 & 2240 & 1216 & 1536 \\
\end{array}
\right)
\ee

\subsection{$G_0$ and $\gamma_0$ in the diagonal canonical basis}

A major implication (section \ref{3}) of eq. \eqref{2.35} for the Hermitian operators in eq. \eqref{dim8set} is that an operator basis exists where $G_0$ and $\gamma_0$ are simultaneously diagonal. \par
For reasons of space, we only write down the transformation that diagonalizes simultaneously the planar contributions to  $G_0$ and $\gamma_0$.\par
A simplifying feature of the planar contributions, which follows from the block form of $\gamma_{00}$ and $G_{00}$ in eqs. \eqref{gamma0M} and \eqref{A0M} respectively, is that the mixing between single- and double-trace operators does not occur. \par
We find by Mathematica a transformation (section \ref{3}), $S$, that diagonalizes simultaneously $\gamma_{00}$ and $G_{00}$:
\begin{small}
\be
S = \left(
\begin{array}{cccccccc}
 0 & -\frac{2}{3} & \frac{1}{3} & \frac{5 \sqrt{13}}{6}+3 & 0 & 0 & 0 & 0 \\
 0 & -\frac{2}{3} & \frac{1}{3} & 3-\frac{5 \sqrt{13}}{6} & 0 & 0 & 0 & 0 \\
 -2 & \frac{2}{3} & \frac{2}{3} & \frac{1}{3} & 0 & 0 & 0 & 0 \\
 2 & 0 & 0 & -\frac{1}{2} & 0 & 0 & 0 & 0 \\
 0 & 0 & 0 & 0 & \frac{1}{656} \left(41-3 \sqrt{41}\right) & \frac{7}{2 \sqrt{41}}-\frac{1}{2} & \frac{1}{8}-\frac{3}{8 \sqrt{41}} & \frac{1}{4}-\frac{11}{4 \sqrt{41}} \\
 0 & 0 & 0 & 0 & \frac{101}{16 \sqrt{697}}-\frac{1}{16} & \frac{1}{2}-\frac{13}{2 \sqrt{697}} & -\frac{1}{8}-\frac{75}{8 \sqrt{697}} & \frac{1}{4}+\frac{75}{4 \sqrt{697}} \\
 0 & 0 & 0 & 0 & \frac{1}{656} \left(3 \sqrt{41}+41\right) & -\frac{1}{2}-\frac{7}{2 \sqrt{41}} & \frac{1}{8}+\frac{3}{8 \sqrt{41}} & \frac{1}{4}+\frac{11}{4 \sqrt{41}} \\
 0 & 0 & 0 & 0 & -\frac{1}{16}-\frac{101}{16 \sqrt{697}} & \frac{1}{2}+\frac{13}{2 \sqrt{697}} & \frac{75}{8 \sqrt{697}}-\frac{1}{8} & \frac{1}{4}-\frac{75}{4 \sqrt{697}} \\
\end{array}
\right)
\ee
\end{small}
Hence:
\begin{small}
\be
\label{gDiag}
\gamma'_{00} = \dfrac{1}{(4\pi)^2}\left(
\begin{array}{cccccccc}
 \frac{22}{3} & 0 & 0 & 0 & 0 & 0 & 0 & 0 \\
 0 & \frac{22}{3} & 0 & 0 & 0 & 0 & 0 & 0 \\
 0 & 0 & -\frac{14}{3} & 0 & 0 & 0 & 0 & 0 \\
 0 & 0 & 0 & 0 & 0 & 0 & 0 & 0 \\
 0 & 0 & 0 & 0 & \frac{1}{6} \left(3 \sqrt{41}+5\right) & 0 & 0 & 0 \\
 0 & 0 & 0 & 0 & 0 & \frac{1}{12} \left(\sqrt{697}+13\right) & 0 & 0 \\
 0 & 0 & 0 & 0 & 0 & 0 & \frac{1}{6} \left(-3 \sqrt{41}+5\right) & 0 \\
 0 & 0 & 0 & 0 & 0 & 0 & 0 & \frac{1}{12} \left(-\sqrt{697}+13\right) \\
\end{array}
\right)  
\ee  
\end{small}
and:
\begin{scriptsize}
\be
\label{A00D}
G'_{00} = \dfrac{1}{\pi^8}\left(
\begin{array}{cccccccc}
 1280 \left(18 \sqrt{13}+65\right) & 0 & 0 & 0 & 0 & 0 & 0 & 0 \\
 0 & 1280 \left(65-18 \sqrt{13}\right) & 0 & 0 & 0 & 0 & 0 & 0 \\
 0 & 0 & 1280 & 0 & 0 & 0 & 0 & 0 \\
 0 & 0 & 0 & 1152 & 0 & 0 & 0 & 0 \\
 0 & 0 & 0 & 0 & 144-\frac{816}{\sqrt{41}} & 0 & 0 & 0 \\
 0 & 0 & 0 & 0 & 0 & 432+\frac{7056}{\sqrt{697}} & 0 & 0 \\
 0 & 0 & 0 & 0 & 0 & 0 & 144+\frac{816}{\sqrt{41}} & 0 \\
 0 & 0 & 0 & 0 & 0 & 0 & 0 & 432-\frac{7056}{\sqrt{697}} \\
\end{array}
\right)
\ee
\end{scriptsize}
are diagonal in the new operator basis $\mathcal{O}'=S\mathcal{O}$:
\bea
\mathcal{O}'_1 &=&  -\frac{2}{3}\mathcal{O}_{842}+\frac{1}{3}\mathcal{O}_{843} + \left(3 + \frac{5\sqrt{13}}{6}\right)\mathcal{O}_{844}\nn \\
\mathcal{O}'_2 &=& -\frac{2}{3}\mathcal{O}_{842}+\frac{1}{3}\mathcal{O}_{843} + \left(3 - \frac{5\sqrt{13}}{6}\right)\mathcal{O}_{844} \nn \\
\mathcal{O}'_3 &=& -2\mathcal{O}_{841} + \frac{2}{3}\mathcal{O}_{842} + \frac{2}{3}\mathcal{O}_{843} + \frac{1}{3}\mathcal{O}_{844},\nn \\
\mathcal{O}'_4 &=& 2\mathcal{O}_{841} - \frac{1}{2}\mathcal{O}_{844} \nn \\
\mathcal{O}'_5 &=& \frac{1}{656} \left(41-3 \sqrt{41}\right)\mathcal{O}_{845} +  \left(\frac{7}{2 \sqrt{41}}-\frac{1}{2}\right)\mathcal{O}_{846} +  \left(\frac{1}{8}-\frac{3}{8 \sqrt{41}}\right)\mathcal{O}_{847} \nn \\
&& + \left(\frac{1}{4}-\frac{11}{4 \sqrt{41}}\right)\mathcal{O}_{848} \nn \\
\mathcal{O}'_6 &=& \left(\frac{101}{16 \sqrt{697}}-\frac{1}{16}\right)\mathcal{O}_{845} +  \left(\frac{1}{2}-\frac{13}{2 \sqrt{697}}\right)\mathcal{O}_{846} + \left(-\frac{1}{8}-\frac{75}{8 \sqrt{697}}\right)\mathcal{O}_{847} \nn \\
&& + \left(\frac{1}{4}+\frac{75}{4 \sqrt{697}}\right)\mathcal{O}_{848} \nn \\
\mathcal{O}'_7 & = & \frac{1}{656} \left(3 \sqrt{41}+41\right)\mathcal{O}_{845} +  \left(-\frac{1}{2}-\frac{7}{2 \sqrt{41}}\right)\mathcal{O}_{846} +  \left(\frac{1}{8}+\frac{3}{8 \sqrt{41}}\right)\mathcal{O}_{847} \nn \\
&& +  \left(\frac{1}{4}+\frac{11}{4 \sqrt{41}}\right)\mathcal{O}_{848} \nn \\
\mathcal{O}'_8 & = & \left(-\frac{1}{16}-\frac{101}{16 \sqrt{697}}\right)\mathcal{O}_{845} + \left(\frac{1}{2}+\frac{13}{2 \sqrt{697}}\right)\mathcal{O}_{846} + \left(\frac{75}{8 \sqrt{697}}-\frac{1}{8}\right)\mathcal{O}_{847} \nn \\ 
&& + \left(\frac{1}{4}-\frac{75}{4 \sqrt{697}}\right)\mathcal{O}_{848}
\eea

\subsection{Computation of $r_i$}

Hence, the eigenvalues of $G'_{00}$ are given by eq. \eqref{A00D}:
\begin{small}
\bea
\label{A00E}
G'_{001} & = & \dfrac{1}{\pi^8}1280 \left(18 \sqrt{13}+65\right) \nn \\
G'_{002} & = & \dfrac{1}{\pi^8}1280 \left(65-18 \sqrt{13}\right) \nn \\
G'_{003} & = & \dfrac{1}{\pi^8}1280 \nn \\
G'_{004} & = & \dfrac{1}{\pi^8}1152 \nn \\
G'_{005}& = & \dfrac{1}{\pi^8}\left(144-\frac{816}{\sqrt{41}}\right) \nn \\
G'_{006} & = & \dfrac{1}{\pi^8}\left( 432+\frac{7056}{\sqrt{697}}\right) \nn \\
G'_{007}& = & \dfrac{1}{\pi^8}\left( 144+\frac{816}{\sqrt{41}}\right) \nn \\
G'_{008} & = & \dfrac{1}{\pi^8}\left( 432-\frac{7056}{\sqrt{697}} \right)
\eea
\end{small}
In order to compute the LO-nonplanar contributions, $G'_{02}$ and $\gamma'_{02}$, to $G_0$ and $\gamma_0$ in the diagonal basis, we adopt the following strategy.
Firstly, by means of Mathematica, we obtain an exact expression to all orders of $\frac{1}{N}$ for the eigenvalues of $G_0$ and $\gamma_0$, which for reasons of space we do not report. Secondly, we expand the exact result in powers of $\frac{1}{N}$. \par
According to the general principles of the large-$N$ expansion, the LO-nonplanar contributions to the eigenvalues -- as opposed to the entries of $G_0$ and $\gamma_0$ in the defining basis in eqs. \eqref{gamma0}, \eqref{gamma0M}, \eqref{A0} and \eqref{A0M} -- only are on the order of $\frac{1}{N^2}$: 
\begin{small}
\bea
\label{A22E}
G'_{021} & = & -\dfrac{1}{\pi^8}\frac{128 \left(23996211283253349853325 \sqrt{13}+87171526429617703123189\right)}{81257571090377185773} \nn \\
G'_{022} & = & \dfrac{1}{\pi^8}\frac{128 \left(23996211283253349853325 \sqrt{13}-87171526429617703123189\right)}{81257571090377185773} \nn \\
G'_{023}  & = & -\dfrac{1}{\pi^8}\frac{15781120}{7653} \nn \\
G'_{024} & = & \dfrac{1}{\pi^8}\frac{726912}{313} \nn \\
G'_{025} & = & \dfrac{1}{\pi^8}\frac{48 \left(21503005810853 \sqrt{41}-179803584559127\right)}{10879768244321} \nn \\
G'_{026}  & = & -\dfrac{1}{\pi^8}\frac{144 \left(81108663810864685 \sqrt{697}+2185373648831844591\right)}{56806020789473521} \nn \\
G'_{027}  & = & -\dfrac{1}{\pi^8}\frac{48 \left(21503005810853 \sqrt{41}+179803584559127\right)}{10879768244321} \nn \\
G'_{028}  & = & \dfrac{1}{\pi^8}\frac{144 \left(81108663810864685 \sqrt{697}-2185373648831844591\right)}{56806020789473521}.
\eea
\end{small}
and:
\begin{small}
\bea 
\label{A22E1}
\gamma'_{021} & = &  \dfrac{1}{(4\pi)^2} 20 \nn \\
\gamma'_{022} & = & -  \dfrac{1}{(4\pi)^2} \frac{11}{7} \nn \\
\gamma'_{023} & = & -  \dfrac{1}{(4\pi)^2}40 \nn \\
\gamma'_{024}& = &  \dfrac{1}{(4\pi)^2} \frac{550}{9} \nn \\
\gamma'_{025}& = &  \dfrac{1}{(4\pi)^2}(10 -  \frac{190}{\sqrt{41}}) \nn \\
\gamma'_{026}& = & -  \dfrac{1}{(4\pi)^2}\frac{121 \left(21607-811 \sqrt{697}\right)}{87822} \nn \\
\gamma'_{027}& = &   \dfrac{1}{(4\pi)^2}(10 +  \frac{190}{\sqrt{41}}) \nn \\
\gamma'_{028} & = & -  \dfrac{1}{(4\pi)^2}\frac{121 \left(21607+811 \sqrt{697}\right)}{87822}
\eea
\end{small}
Interestingly, according to eq. \eqref{gDiag}, in the planar theory the differences $\frac{\gamma'_{001}}{\beta_0}-\frac{\gamma'_{004}}{\beta_0}=\frac{\gamma'_{002}}{\beta_0}-\frac{\gamma'_{004}}{\beta_0}=2$ satisfy the resonant condition in eq. \eqref{Res}, with $\beta_0=\frac{1}{(4 \pi)^2} \frac{11}{3}$ in $SU(N)$ YM theory. \par Yet, to the leading nonplanar order, it follows from eqs. \eqref{gDiag} and \eqref{A22E1} that eq. \eqref{Nonres} is actually satisfied, in such a way that, instead, the nonresonant diagonal renormalization scheme exists to the leading nonplanar order (section \ref{3}).\par
Finally, by employing eqs.  \eqref{A22E} and \eqref{A00E} we obtain the ratios, $r_{i}$, between the LO-nonplanar and planar eigenvalues of $G_0$:
\begin{small}
\bea
r_{1} & = & \left(-\frac{785165809905643651219 }{162515142180754371546}+\frac{9333742321650915751277 }{812575710903771857730 \sqrt{13}}\right)\frac{1}{N^2} \nn \\
r_{2} & = & \left( -\frac{785165809905643651219}{162515142180754371546}-\frac{9333742321650915751277}{812575710903771857730 \sqrt{13}}\right)\frac{1}{N^2} \nn \\
r_{3} & = & -\frac{12329}{7653}\frac{1}{N^2} \nn \\
r_{4} & = &  \frac{631}{313}\frac{1}{N^2} \nn \\
r_{5} & = & \left(-\frac{2173245686161}{265360201081}-\frac{5147390284628}{265360201081 \sqrt{41}}\right)\frac{1}{N^2} \nn \\
r_{6} & = &  \left(-\frac{1333572530869403}{163001494374386}-\frac{32290757869709541}{163001494374386 \sqrt{697}}\right)\frac{1}{N^2} \nn \\
r_{7}& = & \left(-\frac{2173245686161}{265360201081}+\frac{5147390284628}{265360201081 \sqrt{41}}\right)\frac{1}{N^2} \nn \\
r_{8}& = &  \left(-\frac{1333572530869403}{163001494374386}+\frac{32290757869709541}{163001494374386 \sqrt{697}}\right)\frac{1}{N^2}
\eea
\end{small}
whose irrationality for most of the eigenvalues shows that the aforementioned scalar operators belong to the real class. \par
Moreover, as $\frac{\gamma'_0}{\beta_0}$ has a nonvanishing LO-nonplanar contribution, the ratios $r_i$ do not make sense (section \ref{30}) beyond the lowest perturbative order for the aforementioned scalar operators.

\section{Gluonic twist-$2$ operators} \label{5}

\subsection{The defining basis}

In Minkowski space-time the bare gluonic Hermitian twist-$2$ operators that are primary conformal operators for the collinear subgroup are defined in the spinorial basis \cite{gluon} by:
\bea \label{p}
\mathcal{O}^{(s)}_{B} = \Tr f_{11}(x)(i\overrightarrow{D}_++i\overleftarrow{D}_+)^{s-2}C^{\frac{5}{2}}_{s-2}\left(\frac{\overrightarrow{D}_+-\overleftarrow{D}_+}{\overrightarrow{D}_++\overleftarrow{D}_+}\right) f_{\dot{1}\dot{1}}(x) 
\eea 
with maximal collinear spin $s \geq 2$. In the vectorial basis \cite{gluon} they read:
\bea
\mathbb{O}^{(s)}_{B} = - \dfrac{1}{2}g_{\perp}^{\mu\nu}\Tr F_{+\mu}(x)  (i\overrightarrow{D}_++i\overleftarrow{D}_+)^{s-2} C^{\frac{5}{2}}_{s-2}\left(\frac{\overrightarrow{D}_+-\overleftarrow{D}_+}{\overrightarrow{D}_++\overleftarrow{D}_+}\right) F_{+\nu}(x) 
\eea
for even $s$, and:
\bea
\tilde{\mathbb{O}}^{(s)}_{B}= - \dfrac{i}{2}\epsilon_{\perp}^{\mu\nu}\Tr F_{+\mu}(x)  (i\overrightarrow{D}_++i\overleftarrow{D}_+)^{s-2}
C^{\frac{5}{2}}_{s-2}\left(\frac{\overrightarrow{D}_+-\overleftarrow{D}_+}{\overrightarrow{D}_++\overleftarrow{D}_+}\right)  F_{+\nu}(x)
\eea
for odd $s$, with $C^{\frac{5}{2}}_{j-1}(x)$ Gegenbauer polynomials \cite{gluon}. 
We also define:
\bea
\mathcal{O}^{(s)}_{Bk} = (i \partial_+)^{s-k}\Tr f_{11}(x)(i\overrightarrow{D}_++i\overleftarrow{D}_+)^{k-2}C^{\frac{5}{2}}_{k-2}\left(\frac{\overrightarrow{D}_+-\overleftarrow{D}_+}{\overrightarrow{D}_++\overleftarrow{D}_+}\right) f_{\dot{1}\dot{1}}(x) 
\eea 
For $s > k \ge 2$ the above operators are conformal descendants \cite{ConfQCD} of the corresponding primary conformal operators with $k=s$ in eq. \eqref{p}. They mix among themselves under renormalization in $SU(N)$ YM theory, in such a way that each primary conformal operator with spin $s$ only mixes \cite{Twist2,ConfQCD} with the derivatives along the direction $p_+$ of primary conformal operators with lower $s$. \par 
As a consequence, for the renormalized operators with collinear spin $s$ we obtain \cite{Twist2,ConfQCD}:
\bea \label{m}
\mathcal{O}^{(s)}_{i}= \sum_{i \geq k \geq 2} Z_{ik} \mathcal{O}^{(s)}_{Bk}
\eea
where the mixing matrix, $Z$, is lower triangular and independent of $s$ \cite{Twist2,ConfQCD}.

\subsection{$\gamma_0$ in the defining basis}

Hence, $\gamma(g)$ is lower triangular in the defining basis. 
Moreover, it turns out  \cite{Twist2,ConfQCD} that $\gamma_0$ is diagonal and $\gamma_1$ is lower triangular. 
In our notation the eigenvalues of $\gamma_0$ are given \cite{Twist2} by:
\bea \label{an1}
\gamma_{0i}= \frac{2}{(4 \pi)^2} \left(4 \psi(i+1) - 4 \psi(1) -\beta_0 - 8 \frac{i^2+i+1}{(i-1)i(i+1)(i+2)}\right)
\eea
for even $i \geq 2$, and:
\bea \label{an2}
\gamma_{0i}= \frac{2}{(4 \pi)^2}  \left(4 \psi(i+1) - 4 \psi(1) -\beta_0 - 8 \frac{i^2+i-2}{(i-1)i(i+1)(i+2)}  \right)
\eea
for odd $i \geq 3$. Consistently, $\gamma_{02}=0$. \par
From the above equations it follows that, for the twist-$2$ operators in $SU(N)$ YM theory, $\gamma_0$ is independent of $N$ and, therefore, has no nonplanar contribution.\par
Moreover, we verify by means of Mathematica that, up to $i,j=10^3$, the differences $\frac{\gamma_{0i}}{\beta_0}-\frac{\gamma_{0j}}{\beta_0}$ satisfy the nonresonant condition in eq. \eqref{Nonres}, in such a way that the nonresonant diagonal renormalization scheme (section \ref{3}) exists for the aforementioned operators.

\subsection{$G_0$ in the defining basis}

The $2$-point correlators in Minkowskian space-time of all the twist-$2$ operators have been computed in \cite{gluon} to the lowest perturbative order:
\begin{align}  \label{conf0}
&\langle \mathcal{O}^{(s_1)}_{Bs_1}(x)  \mathcal{O}^{(s_2)}_{Bs_2}(y)\rangle 
=  \mathcal{C}_{s_1}(x,y) \delta_{s_1 s_2} 
\end{align}
with:
\begin{align} \label{conf00}
\nonumber
\mathcal{C}_{s}(x,y) =&\frac{1}{(4\pi^2)^2} \frac{N^2-1}{4} \frac{2^{2s+2}i^{2s-4}}{(4!)^2}(s+1)^2(s+2)^2
\frac{(x-y)_+^{2s}}{(\rvert x-y\rvert^2-i\epsilon)^{2s+2}}\\\nonumber
&\sum_{k_1 = 0}^{s-2}\sum_{k_2 = 0}^{s-2}{s\choose k_1}{s\choose k_1+2}{s\choose k_2}{s\choose k_2+2}(-1)^{s-k_2+k_1}\\\nonumber
&(s-k_1+k_2)!(s+k_1-k_2)! \\\nonumber
=&\frac{1}{(4\pi^2)^2} \frac{N^2-1}{4}\frac{2^{2s+2}i^{2s-4}}{(4!)^2}(s+1)^2(s+2)^2
(2s)! \frac{(x-y)_+^{2s}}{(\rvert x-y\rvert^2-i\epsilon)^{2s+2}}\\
&\sum_{k_1 = 0}^{s-2}\sum_{k_2 = 0}^{s-2}{s\choose k_1}{s\choose k_1+2}{s\choose k_2}{s\choose k_2+2}(-1)^{k_2+k_1} \frac{1}{{2s\choose k_1+k_2+2}}
\\\nonumber 
= &\frac{1}{(4\pi^2)^2} \frac{N^2-1}{4} \frac{2^{2s+2}}{(4!)^2}(-1)^s(s-1)s(s+1)(s+2)(2s)!
\frac{(x-y)_+^{2s}}{(\rvert x-y\rvert^2-i\epsilon)^{2s+2}}
\end{align}

\subsection{Relation between $G_0$ and $\gamma_0$ in the defining basis}

The corresponding Euclidean correlators have been obtained by analytic continuation \cite{gluon}.
Hence, the defining basis is orthogonal to the lowest order, and $G_0$ is diagonal. Besides, $\gamma_0$ and $G_0$, being both diagonal, obviously commute (section \ref{3}).

\subsection{Computation of $r_i$}

From eqs. \eqref{conf0} and \eqref{conf00}, it follows that:
\bea
r_{i} = -\frac{1}{N^2} = r_{twist-2}
\eea
is the same one for all of the Hermitian twist-$2$ operators in $SU(N)$ YM theory. Moreover, as $\frac{\gamma_0}{\beta_0}$ has no nonplanar contribution and the canonical nonresonant diagonal renormalization scheme exists at least up to $i=10^3$, $r_{twist-2}$ has a nonperturbative meaning (section \ref{30}) for such operators.

\section{Acknowledgments}

We would like to thank J. A. Gracey for making available to us the computation of $\gamma_0$ for the dimension-$8$ scalar operators in \cite{G2} before publication and for correspondence. \par
The second named author also acknowledges the financial support from the European Union Horizon 2020 research and innovation programme: \emph{High precision multi-jet dynamics at the LHC} (grant agreement no. 772009).

\end{document}